\documentclass[prd,preprint,superscriptaddress,showpacs,byrevtex]{revtex4}
\usepackage{bm}
\def\fsl#1{\setbox0=\hbox{$#1$}           
   \dimen0=\wd0                                 
   \setbox1=\hbox{/} \dimen1=\wd1               
   \ifdim\dimen0>\dimen1                        
      \rlap{\hbox to \dimen0{\hfil/\hfil}}      
      #1                                        
   \else                                        
      \rlap{\hbox to \dimen1{\hfil$#1$\hfil}}   
      /                                         
   \fi}                                         %
\newcommand{\be}{\begin{equation}}
\newcommand{\ee}{\end{equation}}
\newcommand{\bea}{\begin{eqnarray}}
\newcommand{\eea}{\end{eqnarray}}
\newcommand{\beq}{\begin{equation}}
\newcommand{\eeq}{\end{equation}}
\newcommand{\beqs}{\begin{eqnarray}}
\newcommand{\eeqs}{\end{eqnarray}}

\begin{document}
\title{ Boundary Surface Term In QCD Is At Finite Distance Due To Confinement of Quarks and Gluons Inside Finite Size Hadron }
\author{Gouranga C Nayak }\thanks{E-Mail: nayakg138@gmail.com}
%
%
\date{\today}
\begin{abstract}
In this paper we show that since we have not observed quarks and gluons outside the hadron due to the confinement of quarks and gluons inside the hadron, the boundary surface term in QCD is at the finite distance which is at the surface of the finite size hadron. Since the boundary surface is at the finite distance we find that the boundary surface term in QCD is non-zero irrespective of the form of the $r$ dependence of the gluon field ${\hat A}_\mu^a(t,r)$ and the $r$ dependence of the quark field ${\hat \psi}_i(t,r)$ where $a=1,...,8$ and $i=1,2,3$ are the color indices. We show that this is consistent with the fact that the cluster decomposition property fails in QCD due to confinement.
\end{abstract}
\maketitle
\pagestyle{plain}

\pagenumbering{arabic}

\section{Introduction}

Many conservation laws of the nature are derived from the first principle by using the Noether's theorem. For example, by using the Noether's theorem, the energy conservation law can be derived from the time translational invariance, the momentum conservation law can be derived from the space translational invariance and the angular momentum conservation law can be derived from the rotational invariance etc. The Noether's theorem is applied to both classical field theory and to the quantum field theory.

In deriving the conservation law from the Noether's theorem one encounters the surface term which is the spatial volume integration of the spatial coordinate derivative of a function, say $f(r)$. If this function $f(r)$ vanishes at the boundary surface of this volume then the surface term is zero which does not contribute to the conserved quantity derived from the Noether's theorem.

Consider for example the application of the Noether's theorem under the time-space translational invariance in classical electrodynamics in the free space which gives the continuity equation \cite{nkjr}
\bea
\partial_\delta T^{\delta \mu}(x) = 0,~~~~~~~~~~~~T^{\nu \delta}(x) =F^{\nu \mu}(x) F_\mu^\delta(x) +\frac{1}{4} g^{\nu \delta} F_{\mu \sigma}(x)F^{\mu \sigma}(x)
\label{nt1}
\eea
where $T^{\mu \nu}(x)$ is the energy-momentum tensor density of the electromagnetic field. From eq. (\ref{nt1}) we find the Poynting's theorem in free space
\bea
\frac{\partial \epsilon}{\partial t} =-{\vec \nabla} \cdot {\vec S},~~~~~~~~~~~\epsilon=\frac{{\vec E}^2(x)+{\vec B}^2(x)}{2},~~~~~~~~~~~{\vec S}={\vec E}(x) \times {\vec B}(x)
\label{nt2}
\eea
where ${\vec E}(x)$ is the electric field and ${\vec B}(x)$ is the magnetic field. Integrating over the volume $\int d^3x$ of the physical system we find from eq. (\ref{nt2}) that
\bea
\frac{dW_{\rm EM}}{dt} =-\int d^3x {\vec \nabla} \cdot [{\vec E}(x) \times {\vec B}(x)]=-\oint dS ~{\hat n} \cdot [{\vec E}(x) \times {\vec B}(x)]
\label{nt3}
\eea
where ${\hat n}$ is the unit normal to the surface $S$ enclosing the volume $V=\int d^3x$ and $W_{\rm EM}$ is the energy of the electromagnetic field given by
\bea
W_{\rm EM}=\int d^3x [\frac{{\vec E}^2(x)+{\vec B}^2(x)}{2}].
\label{nt4}
\eea

If the surface term $\oint dS ~{\hat n} \cdot [{\vec E}(x) \times {\vec B}(x)]$ is zero (non-zero) then the energy $W_{\rm EM}$ of the electromagnetic field in eq. (\ref{nt3}) is conserved (not conserved). In the situations where the electric field (and the magnetic field) falls as $\frac{1}{r^2}$ then the surface term $\oint dS ~{\hat n} \cdot [{\vec E}(x) \times {\vec B}(x)]$ is zero if the surface is at the infinite distance. However, if the surface is at finite distance then the surface term $\oint dS ~{\hat n} \cdot [{\vec E}(x) \times {\vec B}(x)]$ is non-zero irrespective of the form of the $r$ dependence of ${\vec E}(x) \times {\vec B}(x)$. Also if ${\vec E}(x) \times {\vec B}(x)$ does not fall faster than $\frac{1}{r^2}$ then the surface term $\oint dS ~{\hat n} \cdot [{\vec E}(x) \times {\vec B}(x)]$ is non-zero irrespective of whether the surface is at the finite distance or the surface is at the infinite distance.

From eq. (\ref{nt3}) we find
\bea
\frac{dW_T}{dt}=0
\label{nt5}
\eea
where
\bea
W_T=W_{\rm EM}+W_{\rm flux},~~~~~~~~~~~~~~~~~~W_{\rm flux}=\int d^4x {\vec \nabla} \cdot [{\vec E}(x) \times {\vec B}(x)].
\label{nt6}
\eea
In the $\int d^4x$ integration in eq. (\ref{nt6}) the $\int d^3x$ integration is the definite integral and the $\int dt$ integration is the indefinite integral. From eq. (\ref{nt5}) one finds that it is the total energy $W_T$ which is conserved. From eq. (\ref{nt6}) one finds that the total energy $W_T$ includes the energy $W_{\rm EM}$ of the electromagnetic field as given by eq. (\ref{nt4}) plus the energy flux $W_{\rm flux}$ as given by eq. (\ref{nt6}). This is the Poynting's theorem in electrodynamics in free space.

From the above analysis one finds that the energy flux $W_{\rm flux}$ can be zero or non-zero depending on the $r$ dependence of the electric (magnetic) field and depending on whether the boundary surface term is at the finite distance or the boundary surface term is at the infinite distance in a physical situation.

This implies that one cannot always neglect the boundary surface term as it depends on the physical situation.

One can extend the same analysis to QCD. However, in QCD there are further complications. First of all we have not observed quarks and gluons outside the hadron due to confinement of quarks and gluons inside the hadron. This implies that the quark field ${\hat \psi}_i(t,r)$ and the gluon field ${\hat A}_\mu^a(t,r)$ are not present outside the hadron where $i=1,2,3$ and $a=1,...,8$ are the color indices. As we study QCD in terms of the gluon field ${\hat A}_\mu^a(t,r)$ and the quark field ${\hat \psi}_i(t,r)$ we find that the boundary surface term in QCD can not be at a distance outside the hadron. Since the size of the hadron is finite this implies that the boundary surface term in QCD is not at the infinite distance but the boundary surface term in QCD is at the finite distance which is at the surface of the finite size hadron, see eq. (\ref{lqdd}). As the boundary surface term is at the finite distance, the boundary surface term in QCD is non-zero irrespective of the form of the $r$ dependence of the gluon field ${\hat A}_\mu^a(t,r)$ and the $r$ dependence of the quark field ${\hat \psi}_i(t,r)$. This is consistent with the fact that the cluster decomposition property fails in QCD due to confinement.

This is also consistent with the fact that the renormalized coupling constant in pQED exists at zero momentum transfer which is experimentally measured, see section \ref{qed0} but the renormalized coupling constant in pQCD does not exist at zero momentum transfer due to confinement of quarks and gluons inside the finite size hadron, see section \ref{zeroqcd}. This is because the zero momentum transfer $Q=0$ is equivalent to the infinite distance, which implies that although the QED can be studied at infinite distance but the QCD can not be studied at the infinite distance. This means that the boundary surface term in QED can be at the infinite distance but the boundary surface term in QCD can not be at the infinite distance. This is in agreement with the fact that the boundary surface term in QCD is at the finite distance which is at the surface of the finite size hadron.

Hence we find that there is non-zero QCD flux due to confinement of quarks and gluons inside the finite size hadron. This non-zero QCD flux contributes to the corresponding conserved quantity in QCD derived by using the Noether's theorem from the first principle.

The paper is organized as follows. In section II we discuss that the renormalized coupling constant in perturbative QED exists at zero momentum transfer. In section III we discuss that the renormalized coupling constant in perturbative QCD does not exist at zero momentum transfer due to confinement of quarks and gluons inside the finite size hadron. In section IV we discuss that the cluster decomposition property fails in QCD due to confinement. In section V we show that the unphysical QCD hamiltonian operator cannot predict physical energy eigenvalue of the hadron. In section VI we show that the hadronic matrix element of the energy momentum tensor density operator of QCD does not have exponentially falling $e^{-m_\pi r}$ behavior where $m_\pi$ is the mass of the pion. In section VII we find that the boundary surface term in QCD is at finite distance due to confinement of quarks and gluons inside the finite size hadron. Section VIII contains conclusions.

\section{ Renormalized coupling constant in Perturbative QED Exists At Zero Momentum Transfer }\label{qed0}

Using the photon propagator with one loop self energy $\Pi(Q^2)$ one finds that the renormalized QED coupling $\alpha_{\rm QED}(Q^2)$ is given by
\bea
\alpha_{\rm QED}(Q^2) = \alpha [1+ \frac{\alpha}{3\pi} H(\frac{Q^2}{m^2})]
\label{cp1}
\eea
where $Q^2$ is the momentum transfer square, $\alpha=\frac{1}{137.036}$ is the fine structure constant, $m$ is the mass of the electron and
\bea
H(\frac{Q^2}{m^2})=6\int_0^1 dy~y(1-y)~{\rm ln}[1+\frac{Q^2}{m^2}y(1-y)].
\label{cp2}
\eea
Note that since the photons do not directly interact with each other the electron loop contributes to the one loop self energy of the photon.

Using one loop resumed photon propagator
\bea
1+\Pi(Q^2)+\Pi^2(Q^2)+\Pi^3(Q^2)+...=\frac{1}{1-\Pi(Q^2)}
\label{cp3}
\eea
one finds that at the one loop level the renormalized QED coupling $\alpha_{\rm QED}(Q^2)$ is given by
\bea
\alpha_{\rm QED}(Q^2) = \frac{\alpha}{1-\frac{\alpha}{3\pi} H(\frac{Q^2}{m^2})}.
\label{cp4}
\eea

For $Q^2<<< m^2$ we find from eq. (\ref{cp2})
\bea
H(\frac{Q^2}{m^2})\simeq 6\int_0^1 dy~y(1-y)~\frac{Q^2}{m^2}y(1-y)=\frac{1}{5}\frac{Q^2}{m^2},~~~~~~~~~~~~~Q^2<<< m^2
\label{cp6}
\eea
and for $Q^2>>> m^2$ we find from eq. (\ref{cp2})
\bea
H(\frac{Q^2}{m^2})\simeq 6\int_0^1 dy~y(1-y)~{\rm ln}\frac{Q^2}{m^2}={\rm ln}\frac{Q^2}{m^2},~~~~~~~~~~~~~Q^2>>> m^2.
\label{cp7}
\eea
Using eq. (\ref{cp7}) in (\ref{cp4}) we find
\bea
\alpha_{\rm QED}(Q^2) = \frac{\alpha}{1-\frac{\alpha}{3\pi} {\rm ln}\frac{Q^2}{m^2}},~~~~~~~~~~~~~Q^2>>> m^2
\label{cp8}
\eea
which gives
\bea
\alpha_{\rm QED}(Q_{\rm max}^2) =\infty,~~~~~~~~~~~~~~~~~~~~~~~~~Q^2_{\rm max}=m^2e^{\frac{3\pi}{\alpha}}=10^{274}~{\rm GeV}^2.
\label{cp9}
\eea
Note, however, that for the large value of the QED coupling the pQED is not applicable and hence the non-perturbative QED is necessary which can make the $\alpha_{\rm QED}(Q^2=\infty)$ finite.

From eqs. (\ref{cp2}) and (\ref{cp4}) we find
\bea
\alpha_{\rm QED}(Q^2=0)=\alpha
\label{cp5}
\eea
where  $\alpha=\frac{1}{137.036}$ is the fine structure constant which is experimentally measured. Hence we find that the renormalized coupling constant in pQED exists at zero momentum transfer which is experimentally measured. On the other hand, as we will show in section \ref{zeroqcd}, the renormalized coupling constant in pQCD does not exist at zero momentum transfer due to confinement of quarks and gluons inside the finite size hadron.

\section{ Renormalized coupling constant in perturbative QCD Does Not Exist At Zero Momentum Transfer Due To Confinement of Quarks and Gluons Inside Finite Size Hadron }\label{zeroqcd}

Unlike QED where the photons do not directly interact with each other the gluons in QCD interact with each other. Hence there are additional diagrams in QCD in comparison to QED. In QCD at the one loop level the corrections are from 1) quark-gluon vertex, 2) quark self energy $\Sigma(Q^2)$ and 3) gluon self energy $\Pi(Q^2)$ which includes quark loop, gluon loop and ghost loop.

Adding all the one loop diagrams in QCD one finds, similar to eq. (\ref{cp8}) in QED, that at one loop level the renormalized QCD coupling $\alpha_{\rm QCD}(Q^2)$ at large $Q^2$ is given by
\bea
\alpha_{\rm QCD}(Q^2) = \frac{\alpha_{\rm QCD}(\mu^2)}{1+\beta_0\alpha_{\rm QCD}(\mu^2) {\rm ln}\frac{Q^2}{\mu^2}}
\label{cp10}
\eea
which gives
\bea
\alpha_{\rm QCD}(Q^2=\Lambda^2_{\rm QCD}) =\infty,~~~~~~~~~~~~~~~~~~~~~ \Lambda^2_{\rm QCD}=\mu^2e^{-\frac{1}{\beta_0\alpha_{\rm QCD}(\mu^2)}}
\label{cp11a}
\eea
where $\mu$ is the renormalization scale and $\beta_0$ is the beta function in QCD at the one loop level given by $\beta_0 = \frac{33-2N_f}{12\pi}$ with $N_f$ being the number of quark flavors \cite{grl,ptl}. Using $\Lambda^2_{\rm QCD}$ from eq. (\ref{cp11a}) in (\ref{cp10}) we find
\bea
\alpha_{\rm QCD}(Q^2) = \frac{1}{\beta_0 ~{\rm ln}\frac{Q^2}{\Lambda_{\rm QCD}^2}}
\label{cp15}
\eea
where $\Lambda_{\rm QCD}$ is the mass scale in pQCD which is experimentally extracted to be around 200 MeV.

From eq. (\ref{cp15}) one finds that the QCD coupling increases as $Q^2$ decreases and becomes infinity at $Q^2=\Lambda^2_{QCD}$. Although the pQCD (and in particular the eq. (\ref{cp15})) is not applicable for large value of the coupling constant but still the eq. (\ref{cp15}) hints that the coupling constant in pQCD does not exist at the zero momentum transfer $Q=0$ whereas the coupling constant exists in pQED at zero momentum transfer $Q=0$ (see eq. (\ref{cp5})). This is because the smallest value of $Q$ in eq. (\ref{cp15}) is $\Lambda_{\rm QCD}$ which is around 200 MeV which means $Q=0$ is not reached in eq. (\ref{cp15}). This is consistent with the fact that there is confinement in QCD but there is no confinement in QED.

This is also consistent with the fact that the virtual photon at zero momentum transfer $Q=0$ can interact with the electron but the virtual photon at zero momentum transfer $Q=0$ in the deep inelastic scattering (DIS) cannot interact with the quark. This is because the quark is confined inside the hadron which has a finite size, say $R$. Hence the virtual photon has to have $Q^2\ge \frac{1}{R^2}$ in order for the virtual photon to interact with the quark inside the hadron. This agrees with the fact that for zero momentum transfer $Q=0$ the coupling constant in QCD does not exist whereas for zero momentum transfer $Q=0$ the coupling constant in QED exists. It is useful to mention here that for $Q^2\ge \frac{1}{R^2}$ the pQCD is applicable for very large value of $Q^2$ (say for $Q > $ 5 GeV) and the non-perturbative QCD is applicable for small value of $Q^2$ (say for $Q\sim$ 1-5 GeV).

The zero momentum transfer $Q=0$ is equivalent to the infinite distance. This implies that although the QED can be studied at infinite distance but the QCD can not be studied at the infinite distance. This means that the boundary surface term in QED can be at the infinite distance but the boundary surface term in QCD can not be at the infinite distance. This is in agreement with the fact that the boundary surface term in QCD is at the finite distance which is at the surface of the finite size hadron.

\section{ Cluster Decomposition Property Fails In QCD Due To Confinement }\label{clus}

The cluster decomposition property is proved in abelian quantum field theory. However, the cluster decomposition property in QCD is not studied yet. This is because in order to study cluster decomposition property in QCD one has to use non-perturbative QCD because at large distances the pQCD is not applicable. Since the non-perturbative QCD is not solved yet \cite{np1,gsn} one finds that the cluster decomposition property in QCD is not studied yet.

Note that although the cluster decomposition property in QCD is not studied yet, there are physical arguments that the cluster decomposition property fails in QCD due to confinement, see for example \cite{str,rob,low}. In the following we give another argument to show why the cluster decomposition property fails in QCD.

We know that the QCD fields are quark field ${\hat \psi}_i(x)$ and the gluon field ${\hat A}_\mu^a(x)$. We also know that we have not observed quarks and gluons outside the hadron. The quarks and gluons are confined inside the hadron. The size of the hadron is not infinite but the size of the hadron is finite. Since the hadron has finite size one finds that the quark field ${\hat \psi}_i(x)$ and the gluon field ${\hat A}_\mu^a(x)$ do not exist for $x$ larger than the size of the hadron, see eq. (\ref{lqdd}).

This implies that if one wants to study the cluster decomposition property in QCD using the quark field ${\hat \psi}_i(x)$ and the gluon field ${\hat A}_\mu^a(x)$ in QCD then it has to be studied by using the non-perturbative QCD only for $x$ less than the size of the hadron because the quark field ${\hat \psi}_i(x)$ and the gluon field ${\hat A}_\mu^a(x)$ do not exist for $x$ larger than the size of the hadron, see eq. (\ref{lqdd}). Hence the analysis of the cluster decomposition property in the abelian quantum field theory case at the large distance $x\rightarrow \infty$ can not be applied to QCD because the quark field ${\hat \psi}_i(x)$ and the gluon field ${\hat A}_\mu^a(x)$ do not exist for $x$ larger than the size of the hadron due to the confinement of quarks and gluons inside the finite size hadron. This implies that, unlike abelian quantum field theory, the hadronic matrix element of the energy-momentum tensor density operator of QCD does not have exponentially falling $e^{-m_\pi r}$ behavior at large distance $r\rightarrow \infty$ where $m_\pi$ is the mass of the pion, see \ref{falb} for details.

\section{ Unphysical QCD Hamiltonian Operator cannot predict physical energy eigenvalue of the hadron}\label{unph}

Since we have not directly experimentally observed quarks and gluons we find that the QCD hamiltonian $H_{\rm QCD}$ is unphysical. As mentioned earlier the non-perturbative QCD is not solved yet analytically. Due to this reason one uses numerical lattice QCD method to evaluate the full path integration in QCD. Since the lattice QCD does the path integration of the quark and gluon fields (but not of the hadron field) the lattice QCD calculates the vacuum expectation value of the non-perturbative correlation function of the quark and gluon fields in QCD (but cannot calculate the hadronic observable). In order to convert the vacuum expectation of this non-perturbative partonic correlation function to hadronic observable the lattice QCD inserts the complete set of hadron momentum eigenstates
\bea
\sum_P |P><P|=1
\eea
in between partonic operators using the equation \cite{nkl}
\bea
H_{\rm QCD} |P> =E_H|P>
\label{lqa}
\eea
where $E_H$ is the energy eigenvalue of the hadron. Note that since $|P>$ is the momentum eigenstate of the hadron the momentum eigenstate $|P>$ is physical. Similarly since $E_H$ is the energy of the hadron the energy $E_H$ is physical. However, as mentioned above, since the QCD hamiltonian $H_{\rm QCD}$ is the hamiltonian of the quarks and gluons (not of the physical hadron) the QCD hamiltonian $H_{\rm QCD}$ is unphysical even if the QCD hamiltonian $H_{\rm QCD}$ is gauge invariant and color singlet. Note that the hadron momentum eigenstate $|P>$ in this paper is normalized to unity.

Hence one finds that the eq. (\ref{lqa}) is not correct because the left hand side of eq. (\ref{lqa}) is unphysical but the right hand side of eq. (\ref{lqa}) is physical. In eq. (\ref{lqa}) the QCD hamiltonian $H_{\rm QCD}$ includes all the quarks plus antiquarks plus gluons inside the hadron. Recently we have shown that the total energy $E_{\rm QCD}(t)$ of quarks plus antiquarks plus gluons inside the hadron is given by \cite{nkl}
\bea
E_{\rm QCD}(t) =<P|\int d^3r {\hat T}^{00}(t,r)|P> =  <P|H_{\rm QCD}|P> \neq E_H.
\label{lqc}
\eea
The inequality in the right hand side of eq. (\ref{lqc}) is due to the non-zero energy flux $E_{\rm flux}(t)$ in QCD which arises due to the non-vanishing boundary surface term in QCD because of confinement of quarks and gluons inside the finite size hadron. In eq. (\ref{lqc}) the ${\hat T}^{00}(t,r)$ is the $00$ component of the energy-momentum tensor density operator ${\hat T}^{\lambda \delta}(x)$ in QCD given by
\bea
&& {\hat T}^{\lambda \delta}_{\rm QCD}(x)=\sum_q \frac{i}{2} {\hat {\bar {\hat \psi}}}_l(x)[\gamma^\lambda (\delta_{lj}{\overrightarrow \partial}^\delta-igT^c_{lj}{\hat A}^{\delta c}(x)) -\gamma^\lambda (\delta_{lj}{\overleftarrow \partial}^\delta+igT^c_{lj}{\hat A}^{\delta c}(x))]{\hat {\hat \psi}}_j(x)\nonumber \\
&&+\sum_g {\hat F}^{\lambda \sigma a}(x){\hat F}_\sigma^{\delta a}(x) +\sum_g \frac{1}{4}g^{\lambda \delta} {\hat F}_{\mu \sigma}^a(x){\hat F}^{\mu \sigma a}(x)+(antiquarks)
\label{tld}
\eea
where ${\hat {\hat \psi}}_i(x)$ is the quark field operator, ${\hat A}_\mu^a(x)$ is the (quantum) gluon field operator and
\bea
{\hat F}_{\mu \sigma}^c(x)=\partial_\mu {\hat A}_\sigma^c(x) - \partial_\sigma {\hat A}_\mu^c(x) + gf^{cdb} {\hat A}_\mu^d(x){\hat A}_\sigma^b(x).
\eea
In eq. (\ref{tld}) the QCD energy-momentum tensor density operator ${\hat T}^{\lambda \delta}_{\rm QCD}(x)$ includes all the quarks plus antiquarks plus gluons inside the hadron where the term $(antiquarks)$ means the expression for the antiquarks similar to the expression in the first term for the quarks. From eq. (\ref{lqc}) we find
\bea
H_{\rm QCD} |P> \neq E_H|P>
\label{lqb}
\eea
which does not agree with eq. (\ref{lqa}).

Hence we find that it is not possible to extract the properties of hadron by inserting the complete set of hadron states $\sum_P |P><P|=1$ in between partonic operators in the vacuum expectation of the non-perturbative correlation function of the quark and gluon fields in QCD.

The eq. (\ref{lqb}) will be used in section \ref{falb} to show that the hadronic matrix element of the energy-momentum tensor density operator of QCD does not have exponentially falling behavior $e^{-m_\pi r}$ where $m_\pi$ is the mass of the pion.

\section{ Hadronic Matrix Element of The Energy-Momentum Tensor Density Operator of QCD Does Not Have Exponentially Falling $e^{-m_\pi r}$ Behavior}\label{falb}

Let us consider the momentum conservation equation of partons inside the hadron from the Noether's theorem in QCD given by \cite{nkp}
\bea
\frac{d}{dt} <P|{\hat p}^\mu |P> = \frac{\partial }{\partial t} <P|\int d^3x {\hat T}^{0\mu}_{\rm QCD}(t,{\vec x}) |P>= - \int d^3x \frac{\partial }{\partial x^i} <P| {\hat T}^{i\mu}_{\rm QCD}(t,{\vec x}) |P>
\label{lqd}
\eea
where ${\hat p}^\mu$ is the momentum operator of all the quarks plus antiquarks plus gluons inside the hadron. From eq. (\ref{lqd}) we find that the momentum density operator ${\hat {\cal P}}^\mu(x)$ of the partons inside the hadron is defined as
\bea
{\hat {\cal P}}^\mu(x)={\hat T}^{0\mu}_{\rm QCD}(x)
\label{lqda}
\eea
which by using eq. (\ref{tld}) gives
\bea
&&{\hat {\cal P}}^\mu(x)=\sum_q \frac{i}{2} {\hat {\bar {\hat \psi}}}_l(x)[\gamma^0 (\delta_{lj}{\overrightarrow \partial}^\mu-igT^c_{lj}{\hat A}^{\mu c}(x)) -\gamma^0 (\delta_{lj}{\overleftarrow \partial}^\mu+igT^c_{lj}{\hat A}^{\mu c}(x))]{\hat {\hat \psi}}_j(x)\nonumber \\
&&+\sum_g {\hat F}^{0 \sigma a}(x){\hat F}_\sigma^{\mu a}(x) +\sum_g \frac{1}{4}g^{0 \mu} {\hat F}_{\nu \sigma}^a(x){\hat F}^{\nu \sigma a}(x)+(antiquarks).
\label{lqdb}
\eea
Since the gluon field ${\hat A}_\mu^a(x)$ and the quark field ${\hat \psi}_i(x)$ do not exist outside the hadron due to the confinement of quarks and gluons inside the finite size hadron one finds that the momentum density operator ${\hat {\cal P}}^\mu(x)$ of the partons inside the hadron exists only inside the hadron. Since the momentum density operator ${\hat {\cal P}}^\mu(x)$ of the partons inside the hadron exits only inside the hadron the momentum ${\hat p}^\mu$ of the partons inside the hadron is obtained by integrating over the volume of the hadron
\bea
{\hat p}^\mu =\int_V d^3x {\hat {\cal P}}^\mu(t,{\vec x})
\label{lqdc}
\eea
where $V$ is the volume of the hadron. Hence we find from eqs. (\ref{lqdc}) and (\ref{lqdb}) that the volume integral in eq. (\ref{lqd}) is given by
\bea
\int d^3x =V,~~~~~~~~~~~~~~~~~~~{\rm V~=~Finite~~Volume~~of~~Hadron}
\label{lqdd}
\eea
where $V$ is the volume of the finite size hadron.

This is consistent with section \ref{clus} where we showed that the volume of integration $\int d^3x$ in eq. (\ref{lqd}) can not extend beyond the size of the hadron because in the ${\hat T}^{\mu \nu}_{\rm QCD}(x)$ [see eq. (\ref{tld})] the quark field ${\hat \psi}_i(x)$ and the gluon field ${\hat A}_\mu^a(x)$ do not exist outside the hadron due to the confinement of quarks and gluons inside the finite size hadron.

Also as shown in section \ref{clus} since the cluster decomposition property fails in QCD due to confinement of quarks and gluons inside the finite size hadron one finds that the hadronic matrix element $<P| {\hat T}^{i\mu}_{\rm QCD}(t,r) |P>$ of the QCD energy-momentum density operator does not have an exponentially falling behavior $e^{-m r}$ where $m$ is the lowest relevant mass in the spectrum of the QCD hamiltonian $H_{\rm QCD}$.

Note that although the lowest mass in the hadronic spectrum is the pion but the lowest relevant mass $m$ in the spectrum of the QCD hamiltonian $H_{\rm QCD}$ can not be equal to the pion mass $m_\pi$ because of eq. (\ref{lqb}). This implies that
\bea
e^{-mr} \neq e^{-m_\pi r}.
\label{lqe}
\eea
Hence we find that the hadronic matrix element $<P| {\hat T}^{i\mu}_{\rm QCD}(t,r) |P>$ of the energy-momentum tensor density operator of QCD does not have exponentially falling $e^{-m_\pi r}$ behavior where $m_\pi$ is the mass of the pion.

\section{ Boundary Surface Term In QCD Is At Finite Distance Due To Confinement Of Quarks and Gluons Inside Finite Size Hadron}\label{fd}

To show that the surface term vanishes, one may appeal to the cluster decomposition property in abelian quantum field theory to show that at asymptotically large distance $r$ the $<P|{\hat T}^{i\mu}_{\rm QCD}(t,r)|P>$ has an exponential decay behavior. One may further argue that this decay is at most of order $e^{-mr}$ times a power of $r$ (similar to the cluster decomposition property of abelian quantum field theory) using $m$ as the lowest relevant mass in the spectrum of the QCD Hamiltonian $H_{\rm QCD}$ by assuming that the cluster decomposition property of abelian quantum field theory is valid for QCD. Then one may argue that the lowest mass particle in QCD is pion to claim that the asymptotic behavior of the hadronic matrix element $<P|{\hat T}^{i\mu}_{\rm QCD}(t,r)|P>$ of the energy momentum tensor density operator ${\hat T}^{i\mu}_{\rm QCD}(t,r)$ of QCD has an exponentially falling $e^{-m_\pi r}$ behavior which vanishes as $r\rightarrow \infty$. In this way one may claim that the boundary surface term in QCD vanishes. This argument, however, is not correct which we will show below.

First of all the lowest mass in QCD is not pion but the lowest mass particle in QCD is the up quark. The pion is the lowest mass particle in the hadron spectrum. Hence one should not assume that the mass $m$ in $e^{-mr}$ is $m=m_\pi$ because, as shown in sections \ref{unph} and \ref{falb}, the QCD hamiltonian $H_{\rm QCD}$ is unphysical and hence the unphysical QCD hamiltonian $H_{\rm QCD}$ operating on the physical momentum eigenvector $|P>$ of the hadron can not give the physical energy eigenvalue $E_H$ of the hadron as shown in eq. (\ref{lqb}). Hence one finds that $e^{-mr} \neq e^{-m_\pi r}$, see eq. (\ref{lqe}).

Secondly, applying the cluster decomposition property of abelian quantum field theory to QCD is not correct because, as we have seen in section \ref{clus}, the cluster decomposition property fails in QCD due to confinement where, unlike the $\frac{1}{r}$ potential in abelian quantum field theory, the potential in QCD is an increasing function of distance $r$ causing confinement of quarks and gluons inside the finite size hadron. Hence one can not apply the cluster decomposition property of the abelian quantum field theory to QCD to prove that the $<P|{\hat T}^{i\mu}_{\rm QCD}(t,r)|P>$ of QCD has an exponentially falling $e^{-m_\pi r}$ behavior to vanish at $r\rightarrow \infty$. As mentioned earlier the exact $r$ dependence of the $<P|{\hat T}^{i\mu}_{\rm QCD}(t,r)|P>$ can be calculated by using the non-perturbative QCD which is not solved yet.

Thirdly, since the quarks and gluons do not exist outside the hadron one finds that the volume integral in any QCD calculation of the hadron involving the quark field ${\hat \psi}_i(x)$ and the gluon field ${\hat A}_\mu^a(x)$ can not have a volume larger than the volume of the hadron. Since the volume integration can not be larger than the volume of the hadron one finds that the boundary surface term in any QCD calculation of the hadron involving the quark field ${\hat \psi}_i(x)$ and the gluon field ${\hat A}_\mu^a(x)$ is at the surface of the finite size hadron.

This implies that due to the confinement of quarks and gluons inside the finite size hadron the boundary surface term in QCD is at the finite distance which is at the surface of the finite size hadron, see eq. (\ref{lqdd}).

Since the boundary surface is at the finite distance one finds that the boundary surface term in QCD is non-zero irrespective of the form of the $x$ dependence of the gluon field ${\hat A}_\mu^a(x)$ and the quark field ${\hat \psi}_i(x)$.

\section{Conclusions}
In this paper we have shown that since we have not observed quarks and gluons outside the hadron due to the confinement of quarks and gluons inside the hadron, the boundary surface term in QCD is at the finite distance which is at the surface of the finite size hadron. Since the boundary surface is at the finite distance we have found that the boundary surface term in QCD is non-zero irrespective of the form of the $r$ dependence of the gluon field ${\hat A}_\mu^a(t,r)$ and the $r$ dependence of the quark field ${\hat \psi}_i(t,r)$ where $a=1,...,8$ and $i=1,2,3$ are the color indices. We have shown that this is consistent with the fact that the cluster decomposition property fails in QCD due to confinement.

The boundary surface term at the finite distance due to the confinement of quarks and gluons inside the finite size hadron also plays an important role to study the hadron production from the quark-gluon plasma at RHIC and LHC \cite{nl1,nl2,nl3,nl4,nl5}.

\end{document}